\documentstyle[11pt]{article}
\let\LARGE=\Large
\let\Large=\large
\let\large=\normalsize
\newcommand{\be}[3]{\begin{equation}  \label{#1#2#3}}     
\newcommand{\ee}{ \end{equation}}
\newcommand{\ba}{\begin{array}}
\newcommand{\ea}{\end{array}}
\newcommand{\bea}{\begin{eqnarray}}
\newcommand{\eea}{\end{eqnarray}}
\newcommand{\NP}[3]{{\em Nucl. Phys.}{ \bf B#1#2#3}}

\def\ve{\varepsilon}

\newcommand{\ft}[2]{{\textstyle\frac{#1}{#2}}}

\def\beq{\begin{equation}}
\def\eeq{\end{equation}}
\def\beqa{\begin{eqnarray}}
\def\eeqa{\end{eqnarray}}
\def\F{{\cal F}}

\renewcommand{\d}{\delta}
\newcommand{\pa}{\partial}
\newcommand{\g}{\gamma}

\newcommand{\e}{\epsilon}

\renewcommand{\L}{\Lambda}

\newcommand{\m}{\mu}

\newcommand{\n}{\nu}

\renewcommand{\o}{\omega}
\renewcommand{\O}{\Omega}

\begin{document}
\begin{titlepage}
\begin{center}
\hfill THU-98/44\\
\hfill {\tt hep-th/9812082}\\

\vskip 3cm

{ \LARGE \bf Corrections to macroscopic 
supersymmetric black-hole entropy}

\vskip .3in

{\bf Gabriel Lopes Cardoso$^{1}$\footnote{\mbox{
\tt 
cardoso@phys.uu.nl}}, 
Bernard de Wit$^{1}$\footnote{\mbox{
\tt 
bdewit@phys.uu.nl}} 
and Thomas Mohaupt$^2$\footnote{\mbox{
\tt 
mohaupt@hera1.physik.uni-halle.de}}}
\\

\vskip 1cm


{\em 


\centerline{$^1$Institute for Theoretical Physics, Utrecht University,
3508 TA Utrecht, The Netherlands}

\centerline{$^2$Martin-Luther-Universit\"at Halle-Wittenberg, 
Fachbereich Physik,
D-06099 Halle, Germany}}

\vskip .1in

\end{center}

\vskip .2in

\begin{center} {\bf ABSTRACT } \end{center}
We determine the corrections to the entropy of extremal black 
holes arising from terms quadratic in the Riemann tensor in 
$N=2, D=4$ supergravity theories. We follow Wald's proposal 
to modify the Bekenstein-Hawking area law. The new entropy 
formula, whose value only depends on the electric/magnetic charges, 
is expressed in terms of a single holomorphic function and 
is consistent with electric-magnetic duality. For string effective 
field theories arising from Calabi-Yau compactifications, our result 
for the entropy of a certain class of extremal black-hole solutions 
fully agrees with the counting of microstates performed some time ago 
by Maldacena, Strominger, Witten and by Vafa.


\vfill

December 1998\\
\end{titlepage}

Some time ago the microscopic entropy was determined for certain
black holes arising in Calabi-Yau compactifications of M-theory and 
type-IIA string theory \cite{MSW,Vafa}. The results cannot be 
reconciled with the 
macroscopic entropy formulae based on an effective $N=2$ 
supergravity theory corresponding to tree level in string
perturbation theory \cite{FKS,BCDWKLM} and it was argued that the  
deviations arise from terms in the effective 
action proportional to the square of the Riemann tensor, with 
coefficients linearly related to the second Chern class 
of the Calabi-Yau manifold. The 
latter arise as one-loop corrections in string perturbation 
theory. At the microscopic level  
the deviations correspond to subleading corrections to the 
central charge of a  
two-dimensional conformal field theory and were already present in
the results of \cite{StromVafa}. In \cite{MSW,Vafa} this 
conjectured correspondence between 
microscopic and macroscopic results was verified, based on 
certain assumptions (such as the constancy of the moduli), to 
first order in the coefficients of the $R^2$-terms. Subsequently 
an attempt was made to determine the macroscopic result in the 
context of a full $N=2$ supersymmetric effective field 
theory with $R^2$-terms, exploiting the fixed-point behaviour of 
the moduli fields and many features of special geometry 
\cite{BCDWLMS}. Somewhat surprisingly, however, it was found that 
the first-order correction to the macroscopic entropy  
vanishes, although higher-order terms are indeed present.

In this note we return to this question and completely
resolve the discrepancy. In doing so we are able to go beyond the
specific cases discussed above and to treat a much larger variety of
4-dimensional $N=2$ supersymmetric black-hole solutions. A crucial
ingredient of our work is that we depart from the
Bekenstein-Hawking area formula and follow 
Wald's proposal for the entropy formula based on a Noether charge 
associated with an isometry evaluated at the corresponding  
Killing horizon \cite{Wald}. Let us recall that the black-hole 
solutions that we consider are static, 
rotationally symmetric solitonic interpolations between 
two $N=2$ supersymmetric groundstates: flat Minkowski spacetime at spatial 
infinity and Bertotti-Robinson spacetime at the horizon 
\cite{Gibbons}. Hence we can parametrize the spacetime line element in  
terms of two functions of the radius, 
\beqa
ds^2 = - e^{2g(r)} dt^2 
+ e^{2f(r)} \Big[ dr^2 + r^2 ( \sin^2 \theta \, d\phi^2 +
d \theta^2 ) \Big] \;, 
\label{lineel}
\eeqa
in so-called isotropic coordinates. The 
interpolating solution preserves $N=1$ supersymmetry so that we 
are dealing with a BPS configuration and the black hole is extremal. 
The residual supersymmetry is also responsible for the 
fixed-point behaviour of the moduli and forces them to take 
certain values depending on the electric/magnetic charges
\cite{FKS} (for a recent reference on the fixed-point behaviour, see
\cite{Moore}). 
Assuming fixed-point behaviour, the precise relation follows 
exclusively from electric-magnetic duality considerations 
\cite{BCDWKLM}. By making use of these features, it remains to 
analyze the  
restrictions posed by full $N=2$ supersymmetry on a static 
and spherically symmetric  
geometry. Here it is important to realize that, in view of the 
fact that the action contains $R^2$-terms, one has to make 
optimal use of the existence of a full off-shell 
multiplet calculus for these theories \cite{DWVHVPL}. This enables one to 
perform the analysis in a way that is independent of the 
complications of the action, although ultimately 
the full field equations are indeed satisfied for supersymmetric 
field configuration. 

The multiplet calculus that we employ is based on the 
gauge symmetries of the superconformal algebra. The corresponding high 
degree of symmetry allows for the use of relatively small off-shell 
field representations. One is the Weyl multiplet, whose fields 
comprise the gauge fields corresponding to the superconformal 
symmetries and a  
few auxiliary fields. As we will demonstrate, it is possible to 
analyze the conditions for $N=2$ supersymmetry directly in this 
formulation, postponing a transition to Poincar\'e supergravity 
till the end. Poincar\'e supergravity theories are obtained 
by coupling additional superconformal multiplets containing Yang-Mills and
matter fields to the Weyl multiplet. Under certain conditions the
resulting superconformal theory then becomes  
gauge equivalent to a theory of Poincar\'e supergravity. 
Some of the additional multiplets are necessary to provide
compensating fields and to overcome a deficit in degrees of freedom
between the Weyl multiplet and the 
Poincar\'e supergravity multiplet.  
For instance, the graviphoton, represented by
an abelian vector field in the Poincar\'e supergravity multiplet,
is provided by an $N=2$ superconformal vector multiplet. However,
as mentioned above, the main part of our analysis remains within 
the context of the full superconformal formulation. This implies 
in particular that 
all our results (including (\ref{lineel})) are subject to local 
scale transformations. Only towards the end we will convert to 
expressions that are scale invariant. 

The superconformal algebra contains 
general-coordinate, local Lorentz, dilatation,
special conformal, chiral SU(2) and U(1), supersymmetry 
($Q$) and special supersymmetry ($S$) transformations. The gauge 
fields associated with  
general-coordinate transformations ($e_\m^a$), dilatations ($b_\m$), 
chiral symmetry (${\cal V}^{\;i}_{\m\, j}, A_\m$) 
and $Q$-supersymmetry ($\psi_\m^i$), are realized by independent fields. The 
remaining gauge fields of Lorentz ($\o^{ab}_\m$), special 
conformal ($f^a_\m$) and $S$-supersymmetry transformations
($\phi_\m^i$) are dependent fields.  They are composite objects, which 
depend in a complicated way on the independent fields \cite{DWVHVPL}. 
The corresponding curvatures and covariant fields are contained
in a tensor chiral multiplet, which comprises $24+24$ off-shell 
degrees of freedom; 
in addition to the independent superconformal gauge fields it 
contains three auxiliary fields: a Majorana spinor doublet  
$\chi^i$, a scalar $D$ and a selfdual Lorentz tensor $T_{abij}$ (where $i,
j,\ldots$ are chiral SU(2) spinor indices)\footnote{%
  By an abuse of terminology, 
  $T_{abij}$ is often called the graviphoton field strength.
  It is antisymmetric in  both Lorentz indices $a,b$ and 
  chiral SU(2) indices $i,j$. Its complex conjugate is the anti-selfdual 
  field $T^{ij}_{ab}$. 
  Our conventions are such that SU(2) indices are 
raised and lowered by complex conjugation. The SU(2) gauge field 
${\cal V}_\m^{\;i}{}_j$ is antihermitean and traceless, i.e., 
${\cal V}_\m^{\;i}{}_j+{\cal V}_{\m j}{}^{i}= {\cal 
V}_\m^{\;i}{}_i=0$. }. %
We summarize the transformation rules for some of
the independent fields of the Weyl multiplet
under $Q$- and $S$-super\-symmetry and under 
special conformal transformations, with parameters $\e^i$, $\eta^i$ and
$\L^a_{\rm K}$, respectively,
\bea \d e_\mu{}^a &=&
     \bar{\e}^i\g^a\psi_{\mu i}+{\rm h.c.}\,, \nonumber\\
      \d\psi_\mu^i &=& 2{\cal D}_\mu\e^i
      -\ft18  T^{ab\,ij}\g_{ab}\,\g_\mu\e_j
      -\g_\mu\eta^i \,,\nonumber\\
      \d b_\mu &=&
      \ft12\bar{\e}^i\phi_{\mu i}
      -\ft34\bar{\e}^i\g_\mu\chi_i
      -\ft12\bar{\eta}^i\psi_{\mu i}+{\rm h.c.} +\L_{\rm K}^a\,e_\m^a  
      \,,\nonumber\\   
      \d A_\mu &=& \ft{1}{2}i \bar{\e}^i\phi_{\mu i}
      +\ft{3}{4}i\bar{\e}^i\g_\mu\chi_i
      +\ft{1}{2}i \bar{\eta}^i\psi_{\mu i}+{\rm h.c.}\,, \nonumber\\
      \d T_{ab}^{ ij} &=&
      8 \bar{\e}^{[ i} {R}(Q)_{ab}^{j]}\,, \nonumber\\
      \d\chi^i &=& -\ft1{12}\g_{ab} D\!\llap/\, T^{ab\,ij}\e_j
      +\ft{1}{6} {R}({\cal V})_{ab}{}^{\!i}_{\;j}\,\g^{ab}\e^j
      -\ft{1}{3}i {R}({A})_{ab} \g^{ab} \e^i \nonumber\\
      && +D\,\e^i
      +\ft1{12} T^{ij}_{ab}\g^{ab}\eta_j \,,
 \label{transfo4}
\eea
where ${\cal D}_\mu$ are derivatives covariant with respect to
Lorentz, dilatational, U(1) and SU(2) transformations,
and $D_\mu$ are derivatives covariant with respect to {\it all}
superconformal transformations. Throughout this paper we suppress
terms of higher order in the fermions, as we will be dealing with
a bosonic background. The quantities ${R}(Q)^i_{\m\n}$,  
${R}(A)_{\m\n}$ and ${R}({\cal V})_{\m\n}{}^{\!i}_{\;j}$ 
are supercovariant curvatures 
related to $Q$-supersymmetry, U(1) and SU(2) transformations.

Let us now turn to the abelian vector multiplets, labelled by an 
index $I= 0,1,\ldots,n$. For each value of the index $I$,
there are $8+8$ off-shell degrees of freedom, residing in a 
complex scalar $X^{I}$, a doublet of chiral fermions 
$\Omega_i^{\,I}$, a vector gauge field $W_\mu^{\,I}$,
and a real SU(2) triplet of scalars $Y_{ij}^{\,I}$. 
Under $Q$- and $S$-supersymmetry the fields $X^I$ and $\Omega_i^{\,I}$
 transform as follows:
\bea \d X^{I} &=& \bar{\e}^i\Omega_i^{\,I} \,,\nonumber\\
     \d\Omega_i^{\,I} &=& 2 D\!\llap/\, X^{I}\e_i
     +\ft12 \ve_{ij} \F^{I-}_{\m\n} \g^{\m \n} \e^j
     +Y_{ij}^{\,I}\e^j
     +2X^{I}\eta_i\,.
\label{vrules}\eea
The quantities $\F_{\mu\nu}^I$ are related to the abelian field 
strengths by 
\beq \F_{\mu\nu}^I= F_{\mu\nu}^I
    -\ft14\Big(\ve_{ij}\bar{X}^{I}T^{ij}_{\mu\nu}
+{\rm h.c.}\Big) \;,\quad
F_{\mu\nu}^I=  2\pa_{[\mu}W_{\nu]}^I \;.
\label{calF}
\ee
The covariant quantities of the vector multiplet constitute a 
reduced chiral multiplet. A general chiral multiplet
comprises $16+16$ off-shell degrees of freedom
and carries an arbitrary Weyl weight $w$ (corresponding to the
Weyl weight of its lowest component). The covariant quantities of 
the vector multiplet may be obtained
from a chiral multiplet with $w=1$ by the application of a set of
reducibility conditions, one of which is the Bianchi identity.
Similarly, the covariant quantities of the Weyl multiplet also
constitute a reduced chiral multiplet, denoted by $W^{abij}$,
whose lowest-$\theta$ component is the tensor $T^{abij}$.
{}From this multiplet one may form a scalar (unreduced) chiral
multiplet $W^2 = [W^{abij}\,\varepsilon_{ij}]^2$ which has Weyl and
chiral weights $w=2$ and $c=-2$, respectively \cite{BDRDW}.  

In the following, we will also allow for the presence 
of an arbitrary chiral background
superfield \cite{deWit}, whose component fields will be indicated with a
caret.  Eventually this multiplet will be identified with $W^2$ in
order to generate the $R^2$-terms in the action, but much
of our analysis will not depend on its precise identification. 
We denote its bosonic component fields by ${\hat A}$, ${\hat
B}_{ij}$,  ${\hat F}_{ab}^-$ and by ${\hat C}$.  Here ${\hat A}$ and 
${\hat C}$ denote complex scalar fields, appearing at the 
$\theta^0$- 
and $\theta^4$-level of the chiral background superfield, respectively,
while the symmetric complex SU(2) tensor ${\hat B}_{ij}$ and the
anti-selfdual Lorentz tensor ${\hat F}_{ab}^-$ reside at the 
$\theta^2$-level.  The fermion fields at level $\theta$ and 
$\theta^3$ are denoted by $\hat\Psi_i$ and $\hat\Lambda_i$. Under 
$Q$- and $S$-supersymmetry $\hat A$ and $\Psi_i$
transform as
\beqa
\d {\hat A} &=& \bar{\e}^i {\hat \Psi}_i \,,\nonumber\\
    \d {\hat \Psi}_i
 &=& 2 D\!\llap/\, {\hat A} \e_i
     +\ft12 \ve_{ij} {\hat F}_{ab} \g^{ab} \e^j
     +{\hat B}_{ij} \e^j
     +2 w {\hat A} \eta_i\,,
\eeqa
where $w$ denotes the Weyl weight of the background superfield.
In the presence of this chiral background superfield, the coupling
of the abelian vector multiplets to the Weyl multiplet is encoded in a
function $F(X^I, {\hat A})$, which is holomorphic and
homogenous of degree two,
\beqa
X^I \, F_I + w {\hat A} \, F_{\hat A} = 2 F \;,\quad
F_I = \partial_{X^I} F \;,\quad
F_{\hat A} = \partial_{\hat A} F \;\;\;.
\label{homogeneity}
\eeqa
We will employ this function in what follows, although we are not 
(yet) dealing directly with the action. 

The field equations of the vector multiplets are subject to 
equivalence transformations corresponding to electric-magnetic 
duality, which will not involve the fields of the Weyl multiplet
and of the chiral background. 
As is well-known, two complex $(2n+2)$-component vectors can be 
defined which transform linearly under the SP$(2n+2;{\bf R})$ 
duality group, namely
\beqa
V=\pmatrix{X^I\cr\noalign{\vskip1mm} F_I(X,{\hat A})} \; \quad 
{\mbox{ and}} \quad
\pmatrix{F_{\m\n}^{+I}\cr\noalign{\vskip1mm} G^+_{\m\n\,I}} \;.
\label{seca} 
\eeqa
The first vector has weights $w=1$ and $c=-1$, whereas the second 
one has zero Weyl and chiral weights. 
The field strengths $G^\pm_{\m\n\,I}$ are defined as follows:
\beqa
G^+_{\mu\nu I}={\cal N}_{IJ}F^{+J}_{\mu\nu} + {\cal O}_{\mu\nu 
I}^+\,, \qquad G^-_{\mu\nu I}=\bar {\cal N}_{IJ}F^{-J}_{\mu\nu} + 
{\cal O}_{\mu\nu I}^- \,, \label{defG}
\eeqa
where 
\bea
{\cal N}_{IJ} =\bar F_{IJ}\,, \qquad 
{\cal O}_{\mu\nu I}^+   =\ft14 (F_I-\bar F_{IJ}X^J )\,T_{\mu\nu 
ij}\varepsilon^{ij} +\hat F^+_{\mu\nu} \,\bar  
F_{I{\hat A}} \,. \label{NO} 
\eea
They appear in the field equations of the 
vector fields (in the presence of the background). Solving 
Bianchi identities, ${\cal D}^a ( F^- - F^+)_{ab}^I = 0$, and 
field equations,  ${\cal D}^a (G^- - G^+)_{ab}^I =0$, 
in the static, spherically symmetric geometry (\ref{lineel})
with the chiral background turned on, 
defines the magnetic/electric charges $(p^I,q_J)$, which therefore 
comprise a symplectic vector (note that we use tangent-space 
indices $a=0,1,2,3$ corresponding to $\mu=t,r,\phi,\theta$),
\beqa
F_{01}^{-I} - F_{01}^{+I} = i F_{23}^I = i \,
\frac{{\rm e}^{-2f(r)}}{r^2}\, p^I\;\;,
\nonumber\\
G_{01I}^{-} - G_{01I}^{+} = i G_{23I} = i\, \frac{{\rm 
e}^{-2f(r)}}{r^2} \,q_I \;\;.
\label{charges}
\eea
Assuming a fixed-point behaviour for the moduli at the horizon, 
the symplectic covariance implies that the symplectic vector 
$V$ must be proportional to the symplectic charge vector $(p^I,q_J)$ 
\cite{BCDWLMS}. This yields
\bea
\bar Z \pmatrix{ X^I \,
\cr\noalign{\vskip1mm}  F_J} - Z \pmatrix{ \bar X^I \,
\cr\noalign{\vskip1mm}  \bar F_J} = i \, {\rm e}^{-{\cal K}/2}\; 
\pmatrix{ p^I \cr\noalign{\vskip1mm}  q_J } \;,\quad 
\mbox{with} \quad  Z= {\rm e}^{{\cal K}/2} \,( p^I \, F_I - q_I\, 
X^I)\;,
\label{stab}
\eea
where we introduced the symplectically 
invariant factor (with $w=2$ and $c=0$),
\beq
{\rm e}^{-{\cal K}} = i\Big[\bar X^I\, F_I(X,\hat A) - \bar F_I(\bar X,
\bar{\hat A}) \, X^I \Big] \,,
\label{kaehler}
\eeq
which resembles (but is not equal to) the K\"ahler potential in 
special geometry. By supersymmetry, this factor is related to the 
spinor,  
\bea
\zeta_i \equiv - \Big(\Omega^I_i \,{\pa\over\pa X^I} + \hat\Psi_i \,
{\pa\over\pa \hat A}\Big) {\cal K} 
=-i \,{\rm e}^{\cal K}\Big[  (\bar F_I - \bar 
X^JF_{IJ}) \Omega_i^I  -  \bar X^I F_{I{\hat A}} \, 
\hat\Psi_i\Big]\,.
\eea
It can be shown, using the results contained in \cite{deWit}, 
that $\zeta_i$ is also inert under symplectic  
reparametrizations. 
Under $Q$- and $S$-supersymmetry $\zeta_i$ transforms as 
\bea
\d\zeta_i &=&   - 2 i \,{\rm e}^{\cal K}  (\bar 
F_I D\!\llap/\, X^{I} - \bar X^I  D\!\llap/ \,  F_{I} ) \,
\e_i
-i\, {\rm e}^{\cal K} \Big[(\bar F_I - \bar X^J F_{JI}) Y_{ij}^{\,I} - \bar 
X^IF_{IA} \hat B_{ij}\Big] \,\e^j  \nonumber \\
&&  - \ft{1}{2}i \, 
\ve_{ij} \,
{\cal F}_{ab}^- \, \g^{ab} \e^j +2  \,\eta_i\,,
\label{susyzeta}
\eea
where we ignored higher-order fermionic terms.
Here ${\cal F}_{ab}^-$ denotes the following (tangent-space) 
anti-selfdual tensor,
\beqa
{\cal F}_{ab}^- = {\rm e}^{{\cal K}} \, 
 \left( \bar F_I \,F^{-I}_{ab} - 
\bar X^I \,G_{ab\,I}^- \right) \;\;.
\label{curlyf}
\eeqa

Having fixed the relation between the charges and the values of 
the moduli at the horizon, the next step is to investigate the 
possible horizon geometry by imposing full $N=2$ supersymmetry. 
Let us first observe that the spinor $\zeta_i$ transforms 
inhomogenously under $S$-supersymmetry and hence it can act as 
a compensator for this symmetry. This observation is relevant when 
constructing supersymmetric backgrounds, where one requires
the $Q$-supersymmetry variations of the spinors to vanish modulo a
uniform $S$-transformation. This can conveniently be done by
considering $S$-invariant spinors, constructed by employing
$\zeta_i$. Relevant examples of such spinors are $\O^I_i -
X^I\zeta_i$, $\hat \Psi_i - w \hat A\, \zeta_i$, 
$\chi^i - \ft1{24} T^{ij}_{ab  } \g^{ab}
\zeta_j$ and 
$R(Q)_{ab}^i  
- \ft1{16} T^{cdij} \g_{cd} \g_{ab} \zeta_j$.
Requiring the vanishing of the $Q$-supersymmetry variations of
these $S$-invariant fermionic fields leads to a number of 
stringent conditions on the background, as follows:
\bea
&&\Big(\partial_{\m} - i {\cal A}_{\m} \Big) \,  
\pmatrix{ {\rm e}^{{\cal K}/2} \,
X^I\cr\noalign{\vskip2mm}  {\rm e}^{{\cal K}/2} \, F_I(X,{\hat
A})} = 
\Big(\partial_{\mu} -   i w {\cal A}_{\mu}   \Big) \, \left(
{\rm e}^{w {\cal K}/2} \, 
{\hat A} \right) 
=0\;,\quad
\label{dxf}\\[1.5mm]
&&\F^{-I}_{ab} = -i \, 
 X^I \,   \F^{-}_{ab}  \;, \quad X^I {\hat F}_{ab}^- = w \,
 {\hat A}\,  \F^{-I}_{ab} \;,  
\label{ff}\\[4mm]
&&{\cal D}_c T_{ab}^{ij} =  i \, {\rm e}^{\cal K} \left(\bar X^J
\,{\cal D}_d
F_{J}  - \bar F_J \,{\cal D}_d X^{J} \right) \,
\Big( \d^d_c\, T^{ij}_{ab} 
- 2\d^d_{[a} \,T^{ij}_{b]c} + 2 \d_{c[a}\,T^{ij\,d}_{\,b]}\Big)
\;, \label{dt}
\eea
where ${\cal A}_{\mu}$ denotes a  K\"ahler connection which must
be flat,
\bea
{\cal A}_{\mu} =  \ft{1}{2} \, {\rm e}^{\cal K} \, \left(
 \bar X^J \stackrel{\leftrightarrow}{\pa}_\m  F_{J} 
- \bar F_J \stackrel{\leftrightarrow}{\pa}_\m X^{J}
\right)
\;\;,\qquad  \partial_{[\mu} {\cal A}_{\nu]} =0 
\;\;\;.
\eea
Furthermore we find that the fields $Y^I_{ij}$, $\hat B_{ij}$ and
the curvatures associated with the U(1), SU(2) and dilatational
gauge fields vanish, whereas the field $D$ becomes proportional to 
${\cal F}_{ab}^- \,T^{ij\,ab}$, which will vanish as
we shall prove below. 
Then by using the definition of $G_{\mu \nu I}^-$, the second
relation of (\ref{ff}) and the homogeneity property
(\ref{homogeneity}), the field $T_{ab}^{ij}$ can be
written as follows,
\bea
T^{ij}_{ab}  =
2 i \, \varepsilon^{ij} {\rm e}^{{\cal K}} \, 
\Big[ F_I F^{-I}_{ab}- X^I G_{ab I}^-  \Big]
\,.
\label{tmn}
\eea
This happens to be the same relation that one derives by solving
the field equations for the field $T^{abij}$ (in the presence of
a fixed chiral background), which
incidentally takes the form ${\cal F}_{ab} =0$. Note, however,  
that the right-hand side of this equation also depends on
$T^{abij}$. We also  note the relation
\beq
- \ft14 \varepsilon_{ij} T_{01}^{ij} + i {\cal 
F}^+_{01} = {1\over r^2}\, {\rm e}^{-2f(r)}\,{\rm e}^{{\cal K} /2} \, 
Z\;.
\label{tfz}
\eeq

In addition, we find that the spacetime geometry has to satisfy
\bea
C^-_{ab}{}^{cd} = 
- \ft{1}{24}i \, \varepsilon_{ij}
T^{ef\,ij} \,
{\cal F}^-_{ef} \, 
\Big(\d^{\,cd}_{ab} -\ft12 \varepsilon_{ab}{}^{\!cd}\Big) 
+ \ft{1}8 i \, \varepsilon_{ij}
\Big[ T^{ij}_{ab}\, {\cal F}^{-cd} + 
T^{cd \, ij} \,  {\cal F}^-_{ab}  \Big] \,,
\label{cabcd}
\eea
where the tensor $C^-_{ab}{}^{cd}$, which is antiselfdual in both
pairs of indices, is just the traceless part of the curvature of
the spin connection field $\omega^{ab}_\m$ in the presence of the
dilatational gauge field $b_\m$. Once we fix the scale with
respect to dilatations, this tensor becomes the (antiselfdual
part of the) Weyl tensor. 
As it turns out, in a spherically symmetric geometry, $C^-_{ab}{}^{cd}$ can
only be zero, so that ${\cal F}^-_{ab}\,
T^{cdij}=0$. This proves that the field $D$ is zero indeed. 

At this point two more features are relevant. First of all, as 
alluded to in the introduction, in order to have a consistent
action one needs two compensating supermultiplets \cite{DWVHVPL}.
One of them is a vector multiplet, which we have already taken
into account in our 
analysis given above and which provides the graviphoton. For the
second compensating supermultiplet there are various options,
but they all share the feature that the corresponding spinor does not
transform into a Lorentz vector that takes its values in the
chiral $u(2)$ algebra or into a (anti)selfdual Lorentz tensor. The presence of
the second compensator is also crucial to make the vanishing value of $D$
compatible with the equations of motion. We will not discuss this
rather subtle issue 
in any detail here, but note that the vanishing of the supersymmetry
variation of the spinor belonging to the second compensating multiplet
yields two more conditions, 
\beqa
{\cal F}^-_{ab} =0 \;,\qquad  {\cal A}_\m +A_\m = 0 \;.
\label{cfmn}
\eeqa
Then, it follows from (\ref{dxf}) and (\ref{ff}) that
${\cal D}_\m ( {\rm e}^{{\cal K}/2} X^I) =
{\cal D}_\m ( {\rm e}^{{\cal K}/2} F_I) =
{\cal D}_\m ( {\rm e}^{w {\cal K}/2} {\hat A})=
{\cal F}^{-I}_{ab} = {\hat F}_{ab}^- = 0$.  {}From (\ref{tfz}),
on the other hand, we obtain $
\ft14 \varepsilon_{ij} T_{01}^{ij} = \ft14 i\varepsilon_{ij} T_{23}^{ij} 
= - r^{-2}\, {\rm e}^{-2f(r)}\, {\rm e}^{{\cal K}/2} \, Z$.

Secondly, since the supersymmetry transformations of all fermionic quantities
should vanish in the supersymmetric background,
also derivatives of spinors should have a zero variation. Often
this does not lead to new results, but in the case at hand,
it does.  Namely, requiring
that $D_\m \zeta_i$ vanishes under
$Q$-supersymmetry up to the uniform $S$-transformation that we
mentioned before, yields the following equation for the gauge field
$f_\m^{\,a}$ associated with special conformal boosts:
\beqa
f_{\m a} - \ft12 {\cal D}_\m   {\cal D}_a  {\cal K} - \ft14   {\cal D}_\m 
{\cal K} \,  {\cal D}_a {\cal K}  + \ft18 \left(  {\cal D}_b
{\cal K} \right)^2  \, e_{\m a} = 0 \;\;.
\eeqa
Here, it is important to recall that $f_\m^{\,a}$ is not an independent, but
a composite gauge field \cite{DWVHVPL}, given by  
$f_{\mu}^{\,a} = \ft12 R_{\m}^{\,a} - \ft14 ( D 
+ \ft13 R) e_{\m}^{\,a} + \ft1{16} T^{ab\,ij}\, T_{\m b\,ij}$.

The superconformal theory described above is gauge equivalent to a 
Poincar\'e supergravity theory.  A Poincar\'e frame is chosen
by expressing the results in terms of scale-invariant fields,
such as the rescaled vierbein $e_\m^{\,a}\,\exp[-{\cal K}/2]$,
or, equivalently, by choosing a gauge in which $\cal K$ is constant and 
$b_\m =0$.  In this frame $f_{\m a} = 0$, 
so that the Ricci tensor associated with the spacetime line
element (\ref{lineel}) equals $R_{\m}^{\,a} = -
\ft1{8} T^{ab\,ij}\, T_{\m b\,ij}$. 
Furthermore, using (\ref{cfmn}), we find from 
(\ref{dt}) that ${\cal D}_\m  T^{ab \,ij} =0$.
Now recall that the U(1)-connection $A_\m$ is a flat connection
in the $N=2$ supersymmetric background.  Picking a gauge in which $A_\m=0$ 
and using
the fact that the spacetime  line element  has spherical symmetry, we
conclude that $X^I$,  $F_I$, ${\hat A}$, $T^{ab\,ij}$ and
therefore $Z$  are all constant. From this it follows that
$f(r)+\ln r$ is constant as well. Thus, the only 
nonvanishing components (with tangent-space indices) of the Ricci
tensor are 
$R_{00} =R_{22} =R_{33} = - R_{11} = - \ft1{16} \vert
T^{01\,ij}\varepsilon_{ij}\vert^2$, while the Weyl tensor is zero. 
These requirements restrict the line element (up to diffeomorphisms) to be of
the Bertotti-Robinson type:
\beqa
e^{2g(r)} = e^{- 2f(r)} = \ft1{16} \vert
T^{01\,ij}\varepsilon_{ij}\vert^2 \, r^2  = {\rm e}^{-{\cal K}}
\, {r^2 \over  \vert Z\vert^2 }  \;,
\eeqa
where we note the relation
\beq
 T_{01\,ij} = - 2\,\varepsilon_{ij} {\rm e}^{-{\cal K}/2} \,
Z^{-1} \;.
\label{tzrelation}
\eeq
Hence we have proven that there are two fully supersymmetric 
solutions for off-shell multiplets with an arbitrary chiral
background field that are static and spherically symmetric,
namely Bertotti-Robinson and, when $T^{ab\,ij}=0$,  flat Minkowski spacetime.

Now we write down the Lagrangian of the vector multiplets in the
chiral background \cite{deWit},
\bea
\label{efflag}
8\pi\, {\cal L} &=&  - \ft12 \, {\rm e}^{-{\cal K}} \, R
+ 
\Big[ 
 i {\cal D}^{\mu} F_I \, {\cal D}_{\mu} \bar X^I  
-\ft18i  F_{IJ}\, Y^I_{ij} Y^{Jij} - \ft14 i \hat 
B_{ij}\,F_{{\hat A}I}  Y^{Iij}   \nonumber\\
&&+\ft14 i F_{IJ} (F^{-I}_{ab} -\ft 14 \bar X^I 
T_{ab}^{ij}\varepsilon_{ij})(F^{-J}_{ab} -\ft14 \bar X^J 
T_{ab}^{ij}\varepsilon_{ij})  \nonumber\\
&&-\ft18 i F_I(F^{+I}_{ab} -\ft14  X^I 
T_{abij}\varepsilon^{ij}) T_{abij}\varepsilon^{ij}  
+\ft12 i \hat F^-_{ab}\, F_{{\hat A}I} (F^{-I}_{ab} - \ft14  \bar X^I 
T_{ab}^{ij}\varepsilon_{ij})   \nonumber \\
&&+\ft12 i F_{\hat A}
\hat C -\ft18 i F_{{\hat A}{\hat A}}(\varepsilon^{ik}
\varepsilon^{jl}  \hat B_{ij} 
\hat B_{kl} -2 \hat F^-_{ab}\hat F^-_{ab})
-\ft1{32} iF (T_{abij}\varepsilon^{ij})^2 + {\rm h.c.} \Big]
\;\;.
\eea
Here we have taken into account  the effect from the second
compensator in that we changed the coefficient of the first term
by a factor 3 and we suppressed the terms linear in 
the field $D$. Now we identify the chiral background superfield with 
the superfield $[\varepsilon_{ij} W^{ab\,ij}]^2$, so that $\hat A
= (\varepsilon_{ij} T^{ab\,ij})^2$. It is tedious but
straightforward to determine the higher components of the chiral
background \cite{BDRDW}.  Here we need only the bosonic part
of ${\hat C}$, which is given by ${\hat C} =
- 8 \, 
\varepsilon_{ij}
T^{abij} \{{\cal D}_a , {\cal D}^c \}
T_{cb kl}\varepsilon^{kl} + 16  \,\varepsilon_{ij} T^{abij}
f_a{}^c \, T_{cbkl} \varepsilon^{kl} + 
64\, {\cal R}(M)^-_{cd}{}^{\!ab}\, {\cal 
R}(M)^-_{cd}{}^{\!ab}  + 32\, R({\cal V})^-_{ab}{}^{\!k}{}_l^{~} \, 
R({\cal V})^-_{ab}{}^{\!l}{}_k^{~}$, 
where 
$
{\cal R}(M)_{ab}{}^{cd} = R_{ab}{}^{cd} 
- 4 \d_{[a}{}^{\![c}\, f_{b]}{}^{d]} 
+ \ft1{16} (T^{ijcd}\,T_{ijab} + T^{ij}_{ab}\, T^{cd}_{ij})$.
We note that, after going into the Poincar\'e frame,  the
Lagrangian contains $R^2$-terms, but no terms involving 
derivatives of the Rieman tensor.

The solution we have constructed describes the near-horizon limit
of an extremal black-hole solution in the presence of $R^2$-terms in the
effective Lagrangian (\ref{efflag}).  In order to compute the
macroscopic entropy associated with these solutions, we will use
Wald's formalism \cite{Wald} and view
the entropy as a Noether charge associated with an isometry generated
by the static Killing vector field of the Bertotti-Robinson geometry.
Even though Wald's derivation of the First Law of black-hole mechanics
holds primarily 
for nonextremal black holes, we will in the following assume
that the corresponding entropy formula can be extrapolated to the
case of extremal black holes.  Since our findings are consistent
with electric-magnetic duality, we are confident that this extrapolation
makes perfect sense.  For a class of Lagrangians of the form
(\ref{efflag}) taken in the Poincar\'e frame,
the entropy is, according to Wald's proposal \cite{Wald}, given by
\bea
{\cal S} = \ft1{16}  \,
\oint_{S^2}\; \epsilon_{ab}  \epsilon_{cd}\, 
\frac{\delta (8\pi\,{\cal L})}{\delta R_{abcd}} \;\;, 
\label{waldentr}
\eea
where the indices of the Levi-Civita symbol $\e_{ab}$ run over $a,
b = 0,1$ and where the integral is over a 
spatial cross-section of the Killing horizon, which in this case
is just $S^2$. Using (\ref{lineel}), the integral thus yields an
overall factor $4\pi\, r^2\,\exp [2f(r)]$. Taking the derivative
of the Lagrangian (\ref{efflag}) with respect to the Riemann
tensor in the $(r,t)$-subspace 
leads to various terms. Some of them cancel in the supersymmetric
background and we are left with the variation of the first term
in (\ref{efflag}) and with the variation of the term in $\hat C$
proportional to $f\,T^2$. The result takes the form 
\bea
{\cal S} = \pi\,\Big[ \vert Z\vert^2 -256\,{\rm Im}\,[F_{\hat
A}(X^I ,\hat A)]\,\Big]\;, \quad\mbox{with}\quad {\hat A} =
-64\, \bar Z^{-2}\,{\rm e}^{-{\cal K}} \;\;, 
\label{entropia}
\eea 
where the $X^I$ and therefore $Z$
are specified by (\ref{stab}) in terms of the charges carried by
the extremal black hole. 
The first term in the entropy formula coincides with the
Bekenstein-Hawking area contribution. It is important to note
that the second term involves only the first derivative of the
function $F$ with respect to the background field, which, unlike
higher derivatives, transforms  as a function under symplectic
transformations \cite{deWit}.  Thus our result is consistent with
electric-magnetic duality.

We conclude this letter with the explicit computation of the 
macroscopic entropy
of a certain class of extremal black-hole solutions occuring
in string effective field theories arising from Calabi-Yau compactifications,
and we demonstrate its perfect agreement with the counting of microstates
for this class of black holes performed in \cite{MSW,Vafa}.
We refer to an upcoming publication \cite{CDWM} 
in which we will discuss further
examples, such as heterotic black holes, as well as corrections
to the entropy formulae of \cite{MSW,Vafa} due to the inclusion
of contributions coming from terms in the
Lagrangian proportional to $T^{2(g-1)} R^2$ with constant coupling functions
 \cite{Bershadskyetal,MM,GV}, and also corrections steming 
from certain nonperturbative effects \cite{GV}.

It will be convenient \cite{BCDWKLM} to use rescaled variables
$Y^I = {\rm e}^{{\cal K}/2} {\bar Z} X^I$ and $\Upsilon = {\rm
  e}^{{\cal K}} {\bar Z}^2 {\hat A}$.  Using the homogeneity property of $F$,
it then follows that  
$Z= {\bar Z}^{-1}( p^I F_I( Y , \Upsilon) - q_I Y^I)$ and the
stabilization equations (\ref{stab}) read $Y^I-\bar Y{}^I= i p^I$
and $F_I(Y,\Upsilon) -\bar F_I(\bar Y, \bar \Upsilon) = i
q_I$. Furthermore the entropy formula (\ref{entropia}) takes the form 
${\cal  S} = \pi[ Z\bar Z - 256\, {\rm Im}\, F_{\Upsilon}
(Y,\Upsilon)]$ with $\Upsilon = -64$.
Let us now consider type-IIA string theory compactified on a Calabi-Yau
threefold, in the limit where the volume of the Calabi-Yau threefold
is taken to be large.  For the associated homogenous function
$F(Y,\Upsilon)$ we take (with $I = 0, \dots, n$ and  $A = 1,
\dots, n$) 
\bea
F(Y,\Upsilon) 
= \frac{D_{ABC} Y^A Y^B Y^C}{Y^0} + 
d_{A}\, \frac{Y^A}{Y^0} \; \Upsilon
\;\;,\qquad D_{ABC} = - \ft16 \,  C_{ABC} \;\;,\quad
d_{A} = - \ft{1}{24} \, \ft{1}{64}\; c_{2A} \;\;,
\label{prep2a}
\eea
where the coefficients $C_{ABC}$ denote the intersection numbers
of the four-cycles 
of the Calabi-Yau threefold,  
whereas the coefficients
 ${c_{2A}}$ denote its second Chern-class numbers 
\cite{Bershadskyetal}.  The Lagrangian (\ref{efflag}) associated with
this homogenous function thus contains a term proportional to 
$c_{2A} \, z^A \,R^2$, where $z^A = Y^A/Y^0$.
In the following we will consider
black holes with $p^0 = 0$.  For real $\Upsilon$ the solution of
the stabilization equations 
equals 
\bea
&& Y^0 = {\bar Y}^0 \;,\qquad  (Y^0)^2 = 
\frac{D_{ABC}\, p^A p^B p^C - 4 d_{A}\,
p^A \; \Upsilon}{4 ( q_0 + \ft1{12} D^{AB} q_Aq_B)} \;, \nonumber \\
&& Y^A = \ft16 \, Y^0 \, D^{AB} q_B + \ft{1}{2} i\, p^A \;,\qquad
D_{AB} = D_{ABC}\, p^C \;,\qquad 
D_{AB} D^{BC} = \delta_A^{\,C} \;. 
\eea
Switching off the background field $\Upsilon$ yields the solution presented in
\cite{shma}.  
Furthermore we find
\beq
Z\bar Z = - 4 Y^0 \Big(q_0 + \ft{1}{12} \, D^{AB}q_A q_B\Big) - 2\, {d_A
\,  p^A\over Y^0} \,\Upsilon\,.
\eeq
Inserting this into the entropy formula, and using $\Upsilon = -64$,
yields
\beq
{\cal S} = - 4\pi\, Y^0 \Big(q_0 + \ft{1}{12} \, D^{AB}q_A q_B\Big)\;.
\eeq
Since the coefficients $D_{ABC}$ are negative, 
we take ${\hat q}_0=q_0 + \ft{1}{12} \, D^{AB}q_A q_B < 0$ and
$p^A>0$, so that $(Y^0)^2 > 0$. Choosing $Y^0$ positive as well,
so that the moduli $z^A$ live in the upper half-plane ${\rm Im}\, 
z^A >0$, we thus see that the presence 
of the particular $R^2$-term associated with
(\ref{prep2a}) in the Lagrangian (\ref{efflag}) results
in a correction to the macroscopic entropy which is 
in exact agreement with 
the microscopic entropy formula computed in \cite{MSW,Vafa}:
\bea
{\cal S}_{\rm micro} = 
2 \pi \sqrt{\ft16 \, |{\hat q}_0|  (C_{ABC} \,p^A p^B p^C + 
{c}_{2A} \, p^A) }\;\;.
\eea


{\large \bf Acknowledgements}

\smallskip
\noindent
We thank K. Behrndt, G. Horowitz, T. Jacobson, 
R. Kallosh, B. Kleijn, D. L\"ust, S. Massar and R. Myers
for valuable discussions. 
The work of G.L.C. was supported in part by the European 
Commission TMR 
programme ERBFMRX-CT96-0045.


\end{document}